\documentclass[pra]{revtex4}
\usepackage{amssymb}
\usepackage{amsmath}
\usepackage{graphics}
\newcommand{\vc}[1]{\boldsymbol{#1}}

\begin{document}

\title{ Hermitian and Gauge-Covariant Hamiltonians for a particle in a magnetic field on Cylindrical and Spherical Surfaces}

\author{M. S. Shikakhwa}
\affiliation{ Middle East Technical University Northern Cyprus Campus,Kalkanl\i, G\"{u}zelyurt, via Mersin 10, Turkey}
\author{N. Chair}
\affiliation{Department of physics, The University of Jordan,Amman 11942 Jordan}

\begin{abstract}
We construct the Hermitian Schr\"{o}dinger Hamiltonian of  spin-less   as well as the gauge-covariant Pauli Hamiltonian of  spin one-half particles in a magnetic field that are confined to cylindrical and spherical surfaces. The approach does not require the use of involved differential-geometrical methods and is intuitive and physical, relying on the general requirements of Hermicity and gauge-covariance. The surfaces are embedded in the full three-dimensional space and confinement to the surfaces is achieved by strong radial potentials. We identify the  Hermitian and gauge-covariant (in the presence of a magnetic field) physical radial momentum in each case and set it to zero upon confinement to the surfaces . The resulting surface Hamiltonians are  seen to be automatically Hermitian and gauge-covariant. The well-known geometrical kinetic energy also emerges naturally.
\end{abstract}

\maketitle
\section{introduction}
Writing down the Schr\"{o}dinger equation on a curved surface is a topic that is not covered in  major   undergraduate
\cite{Gaziorowics,Griffiths,Liboff,Shankar,Townsend,Zettili} and graduate \cite{Merzbacher,Sakurai} quantum mechanics textbook, . The reason might be that for a long time,the relevant  quantum systems in curved spaces were mainly  quantum gravity and related topics. These require utilization of the full machinery of quantum field theories in curved space-time, an advanced topic that was - and is still- beyond the scope of any  undergraduate and even graduate physics core course. In recent years, however,the technological advances made possible the synthesis of nano-scale objects with curved surfaces like nanotubes and nanospheres (for reviews see, for example, \cite{CNT,fullerene})  with promising practical applications. In studying the transport properties of charge carriers on these surface, a starting point  in many cases is the single particle Schr\"{o}dinger equation on a curved surface. This last problem is a topic that can be covered in an advanced undergraduate course of quantum mechanics! The present article attempts to set the stage for this. \\
In fact, the most widely accepted scheme for constructing the Schr\"{o}dinger equation on an arbitrary curved surface, mainly known as the thin-layer quantization  was introduced about four decades ago \cite{Koppe,Costa}. In this scheme, one starts in the full 3D Euclidean space, and then the particle is confined to a 2D curved surface embedded in 3D space by a strong confining potential. The extension of this scheme to the important problem of a particle in a magnetic field was done quite recently \cite{Encinosa,ferrari,Jensen1,Jensen2,Ortix1}. Both the original work and its extensions rely on the relatively advanced methodology of tensor analysis, and a knowledge of this as well as classical differential geometry is essential to be able to appreciate the scheme; both of these topics are not usually covered in the standard undergraduate physics curriculum. The scheme that we are going to introduce here is based on a recent work by the authors \cite{shikakhwa and chair1,shikakhwa and chair2}, which is more physical and is  easily accessible to undergraduate physics majors. Although this last scheme is quite general and can be applied to any surface embedded in 3D Orthogonal curvilinear coordinates, we restrict the treatment, again for the sake of clarity and accessibility to larger audiences, to the cases of a cylinder and a sphere. We refer the interested reader to the original works for the discussion of the general case.
\section{Hamiltonian for a spin-less particle on the surface of a cylinder and a sphere}
Let us start by addressing the following seemingly simple question: How can we write down the  Schr\"{o}dinger Hamiltonian for a spin-less particle confined to the cylindrical  surface of an infinitely long cylinder of radius $R$? The straightforward method would be to write down the Schr\"{o}dinger Hamiltonian for a free particle in 3-dimensional space using cylindrical coordinates;
\begin{equation}\label{free H}
H= \frac{-\hbar^2}{2m}\nabla^2=\frac{-\hbar^2}{2m}
\left(\frac{1}{r}\frac{\partial}{\partial r}(r\frac{\partial}{\partial r })+\frac{1}{r^2}\frac{\partial^2}{\partial\theta^2}
+\frac{\partial^2}{\partial z^2}\right)
\end{equation}
then, arguing that since on the surface of a cylinder, the coordinate $r$ will be fixed to the constant value $R$,we drop the derivative with respect to $r$  and set this latter to $R$ in the Hamiltonian to get :

\begin{equation}\label{free H on a cylinder}
H= \frac{-\hbar^2}{2m}\left(\frac{1}{R^2}\frac{\partial^2}{\partial\theta^2}+\frac{\partial^2}{\partial z^2}\right)
\end{equation}
This "pragmatic approach"  is not correct, however, for at least two reasons: The first is that it fails to generate the so-called "geometric kinetic energy" (GKE) , which is a correction for the kinetic energy term that results from the curvature of the cylindrical surface, and the second is that once one introduces other couplings  into the Hamiltonian; an electromagnetic field , for example, this approach gives a non-Hermitian Hamiltonian as we are going to show below. We will introduce now a systematic and physical approach to the problem which is based on the  recent works \cite{shikakhwa and chair1,shikakhwa and chair2} . The idea, is to view the confinement process  as being achieved by a strong radial potential that squeezes the particle  to within a thin layer of width $d$ around the surface and eventually considering the limit $d\rightarrow 0$ pinning the particle to the surface. In doing so, the physical degrees of freedom in the radial direction are suppressed and so, once correctly identified, can be dropped from the Hamiltonian leaving one with the surface Hamiltonian. In this article, we are going to address the important question of identifying the physical radial degrees of freedom, in particular, the radial momentum.\\
To this end, we first recall that the radial momentum is not the simple radial derivative $-i\hbar\frac{\partial}{\partial r}$. This is because this is not Hermitian. Explicitly:
 \begin{eqnarray}\label{Hermicity calculation}
   \langle \Psi|-i\hbar\frac{\partial}{\partial r}\Psi\rangle &=& \int_{0}^{2\pi}d\theta\int_{-L}^{L}dz\int_{R-d/2}^{R+d/2}rdr\Psi^*(\mathbf{r})(-i\hbar\frac{\partial}{\partial r})\Psi(\mathbf{r}) \\\nonumber
    &=&\int_{0}^{2\pi}d\theta\int_{-L}^{L}dz\int_{R-d/2}^{R+d/2}rdr (\frac{i\hbar}{r})\Psi^*(\mathbf{r})\Psi(\mathbf{r})+ \int_{0}^{2\pi}d\theta\int_{-L}^{L}dz\int_{R-d/2}^{R+d/2}rdr(-i\hbar\frac{\partial}{\partial r}\Psi(\mathbf{r}))^*\Psi(\mathbf{r}) \\\nonumber
    &=&\langle -i\hbar(\frac{\partial}{\partial r}+\frac{1}{r})\Psi|\Psi\rangle\neq\langle -i\hbar\frac{\partial}{\partial r}\Psi|\Psi\rangle
 \end{eqnarray}
Here, the radial integral was carried out over the interval $[R-d/2,R+d/2]$ which is the thin-layer of thickness $ d $ about the surface, and the surface term
$ -i\hbar \int_{0}^{2\pi}d\theta\int_{-L}^{L}dz \left[r\Psi^*(\mathbf{r})\Psi(\mathbf{r})\right]_{R-d/2}^{R+d/2} $  that results from the partial integration was dropped as the wave function was assumed to vanish at $R\pm d/2$ which is indeed an admissible boundary condition for a strong symmetric confining potential. As for the limits of the $z$-integration, we have chosen to normalize the wave function over a cylinder symmetric about $z=0$ of arbitrary length $2L$. The idea is similar to the box normalization of the free particle wave function. In fact, the above results actually tell us what could be a Hermitian radial momentum operator. Indeed, it is straightforward, following similar calculations as  above to check that the operator \cite{shikakhwa and chair1,shikakhwa and chair2, Liu1}
\begin{equation}\label{hermitian p}
p_r^{cyl}\equiv -i\hbar(\frac{\partial}{\partial r}+\frac{1}{2r})
\end{equation}
is Hermitian. Note also that $p_r^{cyl}$, being different from $-i\hbar\frac{\partial}{\partial r}$ by a function of $r$ only satisfies the standard canonical commutation relations;
\begin{equation}\label{CCR}
[r,p_r^{cyl}]=i\hbar,\;\;\;\;[p_r^{cyl},p_r^{cyl}]=0
\end{equation}
 and is, therefore, the canonical momentum operator conjugate to $r$. Now, having identified the Hermitian canonical  ( and thus the physical) radial momentum operator, we can express the kinetic energy in terms of it. Noting that,
\begin{equation}\label{laplacian in terms of cyl p}
\frac{-\hbar^2}{2m}\left(\frac{1}{r}\frac{\partial}{\partial r}(r\frac{\partial}{\partial r})\right)=
\frac{(p_r^{cyl})^2}{2m}+\frac{\hbar^2}{8mr^2}
\end{equation}
 the Hamiltonian, Eq.(\ref{free H}), for a free particle can casted in the form:
\begin{equation}\label{free H canonical cylinder}
H= \frac{-\hbar^2}{2m}\nabla^2=\frac{(p_r^{cyl})^2}{2m}-\frac{\hbar^2}{8mr^2}-\frac{\hbar^2}{2m}\left(\frac{1}{r^2}\frac{\partial^2}{\partial\theta^2}
+\frac{\partial^2}{\partial z^2}\right)
\end{equation}
Now, as we have discussed in the paragraph following Eq.(\ref{free H on a cylinder}), a confining radial potential $V(r)$ - not shown explicitly in the above Hamiltonian - pushes the particle to within a layer of thickness $d$ . If the potential is strong enough  we can take the limit $d\rightarrow 0$ thus having the particle pinned to the surface. In this case, the radial degree of freedom is "frozen" and so the radial momentum operator can be dropped from the Hamiltonian, Eq.(\ref{free H canonical cylinder}),leaving us with the surface Hamiltonian:
\begin{equation}\label{hermitian H on a cylinder }
H^{cyl}=\frac{-\hbar^2}{2m}\left(\frac{1}{R^2}\frac{\partial^2}{\partial\theta^2}
+\frac{\partial^2}{\partial z^2}\right)-\frac{\hbar^2}{8mR^2}
\end{equation}
The last term in the above Hamiltonian is the so-called geometric kinetic energy (GKE) mentioned earlier (see Appendix 1 for a discussion of the origins of this term), and it appears naturally here as we work with the physical radial momentum. While there has not been , so far, a direct experimental measurement of this GKE, a recent experiment \cite{optical GKE}  confirmed the existence of such term in optical analogues of quantum systems.   
Note than one can construct the Hamiltonian for a particle on a ring of radius $R$ from the Hamiltonian, Eq.(\ref{hermitian H on a cylinder }),in the same manner: One assumes a squeezing potential $V(z)$ , notes that in this case the Hermitian momentum is just $p_z=-i\hbar\frac{\partial}{\partial z}$ as can be checked easily by proceeding as in Eq.(\ref{Hermicity calculation}) and assuming periodic boundary conditions along the $z$-axis to drop the surface term. The resulting Hermitian Hamiltonian will now read:
\begin{equation}\label{hermitian H on a ring }
H^{ring}=\frac{-\hbar^2}{2m}\left(\frac{1}{R^2}\frac{\partial^2}{\partial\theta^2}
\right)-\frac{\hbar^2}{8mR^2}
\end{equation}
One can apply the same procedure to confine a particle to the surface of a sphere of radius R. Recalling that the Laplacian in spherical coordinates is:
\begin{equation}\label{laplacian spherical}
\nabla^2=\frac{1}{r^2}\frac{\partial}{\partial r}(r^2\frac{\partial}{\partial r})+
\frac{1}{r^2\sin\theta}\frac{\partial}{\partial\theta}(\sin\theta\frac{\partial}{\partial\theta})+\frac{1}{r^2\sin^2\theta}\frac{\partial^2}{\partial\phi^2}
\end{equation}

 One can -following the same procedure as in Eqs.(\ref{Hermicity calculation}) and (\ref{hermitian p})-  check that the Hermitian and at the same time  canonical radial momentum operator in this case is:
\begin{equation}\label{hermitian p sphere}
p_r^{sph}\equiv -i\hbar(\frac{\partial}{\partial r}+\frac{1}{r})
\end{equation}
Thus, we write the analogue of Eq.(\ref{laplacian in terms of cyl p}) as:
\begin{equation}\label{laplacian in terms of sph p}
\frac{-\hbar^2}{2m}\left(\frac{1}{r^2}\frac{\partial}{\partial r}(r^2\frac{\partial}{\partial r})\right)=
\frac{(p_r^{sph})^2}{2m}+0
\end{equation}
We have shown the zero explicitly to emphasize that, unlike the case with the cylinder, the GKE term cancels out in this case (see Appendix 1).  The counterpart of Eq.(\ref{free H canonical cylinder}) is thus:
\begin{equation}\label{free H canonical sphere }
H=\frac{-\hbar^2}{2m}\nabla^2=\frac{-\hbar^2}{2m}\left(\frac{(p_r^{sph})^2}{2m}+\frac{1}{r^2\sin^2\phi}\frac{\partial^2}{\partial\theta^2}+
\frac{1}{r^2\sin\phi}\frac{\partial}{\partial\phi}(\sin\phi\frac{\partial}{\partial\phi})\right)
\end{equation}
The Hermitian Hamiltonian on the surface of a sphere follows readily by dropping the first term which contains the physical radial momentum operator and setting $r$ to $R$  :
\begin{equation}\label{hermitian H sphere }
H^{sph}=\frac{-\hbar^2}{2m}\left(\frac{1}{R^2\sin^2\phi}\frac{\partial^2}{\partial\theta^2}+
\frac{1}{R^2\sin\phi}\frac{\partial}{\partial\phi}(\sin\phi\frac{\partial}{\partial\phi})\right)
\end{equation}
\section{Hermitian Hamiltonians in the presence of magnetic fields}
We now move to the task of constructing Hermitian Hamiltonians on surfaces in the presence of a magnetic field. We consider spin one-half particle to include also the Zeeman term. As we will see, identifying the radial physical degrees of freedom in this case should be done more carefully . We start with a spin one-half particle in an arbitrary magnetic field whose Hamiltonian is given by the Pauli equation:
\begin{equation}\label{Pauli equation}
\frac{(\vc{\sigma}\cdot\vc{\Pi})^2}{2m}=\frac{1}{2m}(\vc{p}-e\vc{A})^2-\frac{e\hbar}{2m}\vc{\sigma}\cdot\vc{B}
\end{equation}
Here, $\sigma$'s are the Pauli spin matrices, $\vc{A}$  the vector potential; $\vc{B}=\vc{\nabla}\times \vc{A}$  the magnetic field and $\vc{\Pi}\equiv (\vc{p}-e\vc{A})$ is the kinematical momentum.
In cylindrical coordinates, the above Hamiltonian expands as :
\begin{eqnarray}\label{H cylinder}
H&=& \frac{-\hbar^2}{2m}\nabla^2+ i\frac{\hbar e}{m}\left(A_{r}\frac{\partial}{\partial r}+\frac{1}{r}A_{\theta}\frac{\partial}{\partial\theta}
+A_{z}\frac{\partial}{\partial z}\right)+i\frac{\hbar e}{2m}\left(\frac{1}{r}\frac{\partial}{\partial r}(rA_{r})+\frac{1}{r}\frac{\partial}{\partial\theta}(A_{\theta})
+\frac{\partial A_{z}}{\partial z}\right)\nonumber\\&+&\frac{ e^2}{2m}\vc{A}.\vc{A}-\frac{e\hbar }{2m}\vc{\sigma}\cdot\vc{B}
\end{eqnarray}
where the Laplacian in cylindrical coordinates is given by  Eq.(\ref{free H}).

 We first demonstrate the fact mentioned in section one, that if we follow the "pragmatic approach" in constructing the Hamiltonian on the surface of a cylinder by setting $r$ to $R$, and dropping the derivative with respect to $r$ from the Hamiltonian, Eq. (\ref{H cylinder}), then we get a non-Hermitian Hamiltonian. To this end, we have the Hamiltonian, Eq.(\ref{H cylinder}), upon setting $\frac{\partial}{\partial r} $ to zero and $r$ to $R$ as:
\begin{eqnarray}\label{H' cylinder}
H^{\prime} &=& \frac{-\hbar^2}{2m}\nabla'^2+ i\frac{\hbar e}{m}\left(\frac{1}{R}A_{\theta}\frac{\partial}{\partial\theta}
+A_{z}\frac{\partial}{\partial z}\right)+i\frac{\hbar e}{2m}\left(\frac{A_{r}}{R}+\frac{\partial}{\partial r}(A_{r})+\frac{1}{R}\frac{\partial}{\partial\theta}(A_{\theta})
+\frac{\partial A_{z}}{\partial z}\right)\nonumber\\&+&\frac{ e^2}{2m}\vc{A}.\vc{A}-\frac{e\hbar }{2m}\vc{\sigma}\cdot\vc{B}
\end{eqnarray}
where $\nabla'^2 $ is constructed from the Laplacian in Eq. (\ref{free H}) with $\frac{\partial}{\partial r} $ set to zero and $r$ to $R$, i.e, it is the " Laplacian at the surface ". Since it is well-known that the Laplacian operator is Hermitian , $ \vc{A}$  is real and the Pauli matrices are also Hermitian, we need to focus only on the second and third terms to demonstrate non-Hermicity. The third term is just a real function multiplied by $i$, and so it is anti-Hermitian,
\begin{eqnarray}\label{3rd term}
 \langle \Phi|i\frac{\hbar e}{2m}\left(\frac{A_{r}}{R}+\frac{\partial}{\partial r}(A_{r})+\frac{1}{R}\frac{\partial}{\partial\theta}(A_{\theta})
+\frac{\partial A_{z}}{\partial z}\right)|\Phi\rangle & =&\langle -i\frac{\hbar e}{2m}\left(\frac{A_{r}}{R}+\frac{\partial}{\partial r}(A_{r})+\frac{1}{R}\frac{\partial}{\partial\theta}(A_{\theta})
+\frac{\partial A_{z}}{\partial z} \right)\Phi|\Phi\rangle
\end{eqnarray}
Here, it is important to keep in mind that $\Phi=\Phi(\theta, z, r=R)$ is the wave function at the surface, normalized as $\int_{-L}^{L}dz \int_ {0}^{2\pi}Rd\theta \Phi^{*}\Phi=1 $. Let us examine the Hermicity of the second  term now:
\begin{eqnarray}\label{2nd term}
\langle \Phi| \frac{i\hbar e}{m}\left(\frac{1}{R}A_{\theta}\frac{\partial}{\partial\theta}+A_{z}\frac{\partial}{\partial z}\right)|\Phi\rangle &=&\int_{-L}^{L}dz\int_ {0}^{2\pi} Rd\theta \left(\Phi^{*}(\frac{i\hbar e}{m})\left(\frac{1}{R}A_{\theta}\frac{\partial}{\partial\theta}+A_{z}\frac{\partial}{\partial z}\right)\Phi\right)\nonumber\\&=&\int_{-L}^{L}dz\int_ {0}^{2\pi} Rd\theta \left[ \left((\frac{i\hbar e}{m})\left(\frac{1}{R}\frac{\partial A_{\theta}}{\partial\theta}+\frac{\partial A_{z}}{\partial z}\right)\Phi\right)^{*}\Phi  \right]\nonumber\\ &+&\int_{-L}^{L}dz\int_ {0}^{2\pi} Rd\theta \left[ \left((\frac{i\hbar e}{m})\left(\frac{1}{R}A_{\theta}\frac{\partial}{\partial\theta}+A_{z}\frac{\partial}{\partial z}\right)\Phi\right)^{*}\Phi  \right]\nonumber\\&=&\langle (\frac{i\hbar e}{m})\left(\frac{1}{R}\frac{\partial A_{\theta}}{\partial\theta}+\frac{\partial A_{z}}{\partial z}\right)\Phi|\Phi\rangle+\langle \frac{i\hbar e}{m} \left(\frac{1}{R}A_{\theta}\frac{\partial}{\partial\theta}+A_{z}\frac{\partial}{\partial z}\right)\Phi|\Phi\rangle
\end{eqnarray}
Therefore, Eqs. (\ref{3rd term}) and (\ref{2nd term}) give:
\begin{eqnarray}\label{2nd and 3rd terms}
&&\langle \Phi| \frac{i\hbar e}{m}\left(\frac{1}{R}A_{\theta}\frac{\partial}{\partial\theta}+A_{z}\frac{\partial}{\partial z}\right)+i\frac{\hbar e}{2m}\left(\frac{A_{r}}{R}+\frac{\partial}{\partial r}(A_{r})+\frac{1}{R}\frac{\partial}{\partial\theta}(A_{\theta})
+\frac{\partial A_{z}}{\partial z}\right)|\Phi\rangle\nonumber\\& =&\langle\left[ \frac{i\hbar e}{m} \left(\frac{1}{R}A_{\theta}\frac{\partial}{\partial\theta}+A_{z}\frac{\partial}{\partial z}\right)+i\frac{\hbar e}{2m}\left(\frac{1}{R}\frac{\partial}{\partial\theta}(A_{\theta})
+\frac{\partial A_{z}}{\partial z}) \right)\right]\Phi|\Phi\rangle+\langle(-i\frac{\hbar e}{2m})\left(\frac{A_{r}}{R}+\frac{\partial}{\partial r}(A_{r})\right)\Phi|\Phi\rangle
\end{eqnarray}
Note the minus sign in the last term on the RHS of the above equation which brakes Hermicity. Having demonstrated the violation of the Hermicity of the surface Hamiltonian constructed by the "pragmatic approach" we move now to show how to construct the right one.

 At this stage, we also stress that there is another condition, in addition to Hermicity,that the surface Hamiltonian is to satisfy, which is gauge-covariance. The observation of this condition demand that the quantity we should take as the physical radial momentum should be both Hermitian  and gauge-covariant. The quantity that satisfies this  is the radial component of the kinematical momentum;
 \begin{equation}\label{k momentum}
 \Pi_{r}=p_{r}-eA_{r}
 \end{equation}
 rather than the canonical radial momentum $p_{r}$. The kinematical momentum $ \vc{\Pi}=\vc{p}-e\vc{A}$ is known to be gauge-covariant (see the appendix) and $\dot {{\vc{r}}}=\frac{d\vc{r}}{dt}=\frac{[\vc{r},H]}{i\hbar}=\vc{\Pi}.$
 Note that ${p_{r}}$ in Eq. (\ref{k momentum}) is the Hermitian canonical momentum $p_r^{cyl}$ defined in Eq.(\ref{hermitian p})

 We now proceed along the same lines as in the previous section . We first note that
 \begin{eqnarray}\label{(k momentum)^2}
 {\Pi_{r}}^2 &=&\frac{-\hbar^2}{2m}\left(\frac{1}{r}\frac{\partial}{\partial r}(r\frac{\partial}{\partial r})\right)+\frac{{\hbar}^2}{8mr^2}+\frac{i\hbar e}{m}(A_{r}\frac{\partial}{\partial r})\nonumber \\&+&\frac{i\hbar e}{2m}\left(\frac{A_{r}}{R}+\frac{\partial}{\partial r}(A_{r})\right)+ \frac{e^2}{2m}{A_{r}}^2
 \end{eqnarray}
 So, expressing the Hamiltonian, Eq. (\ref{H cylinder}),in terms of ${\Pi_{r}}^2$, we have it as:
 \begin{eqnarray}\label{(H cylinder prime}
 H&=&\frac{ {\Pi_{r}}^2}{2m}-\frac{{\hbar}^2}{8mr^2}-\frac{\hbar^2}{2m}\left(\frac{1}{r^2}\frac{\partial^2}{\partial\theta^2}
+\frac{\partial^2}{\partial z^2}\right)\nonumber\\&+&(\frac{i\hbar e}{m})(\frac{A_{\theta}}{r}\frac{\partial}{\partial \theta}+A_{z}\frac{\partial}{\partial z})+ (\frac{i\hbar e}{2m})(\frac{1}{r}\frac{\partial A_{\theta}}{\partial \theta}+ \frac{\partial A_{z}}{\partial z})+\frac{ e^2}{2m}({A_{\theta}}^2+{A_{z}}^2)
\end{eqnarray}
Note  the emergence of GKE term here, too, and that not only the radial derivatives, but also all reference to $A_r$ is now contained in $\frac{ {\Pi_{r}}^2}{2m}$.

Now, upon squeezing particle the surface of the cylinder by the deep potential $V(r)$, we can, in the limit when $ r \rightarrow R$ ($d  \rightarrow 0$), drop the radial physical momentum $ \Pi_{r}$ from the Hamiltonian Eq. (\ref{(H cylinder prime}), and set $r$ to $R$ to get the surface Hamiltonian as :
\begin{eqnarray}\label{(H cylinder prime}
 H^{cyl}&=&\frac{-\hbar^2}{2m}\left(\frac{1}{R^2}\frac{\partial^2}{\partial\theta^2}
+\frac{\partial^2}{\partial z^2}\right)-\frac{{\hbar}^2}{8mR^2}\nonumber\\&+&(\frac{i\hbar e}{m})(\frac{A_{\theta}}{r}\frac{\partial}{\partial \theta}+A_{z}\frac{\partial}{\partial z})+ (\frac{i\hbar e}{2m})(\frac{1}{r}\frac{\partial A_{\theta}}{\partial \theta}+ \frac{\partial A_{z}}{\partial z})+\frac{ e^2}{2m}({A_{\theta}}^2+{A_{z}}^2)
\end{eqnarray}
The problematic terms violating Hermiticity (see Eq. (\ref{2nd and 3rd terms})) have now  dropped all the way along with $ \Pi_{r}$, leaving a Hermitian Hamiltonian. To see that this Hamiltonian is also gauge-covariant, we cast it in terms of the covariant derivatives.  To this end, define the on-the-surface covariant derivative ${\vc{D}}^{\prime}$:
 \begin{equation}\label{cov derivative}
 {\vc{D}}^{\prime}=-i\hbar\left({\vc{\nabla}}^{\prime}-\frac{i e}{\hbar}\vc {A}^{\prime}\right)
 \end{equation}
 with components
\begin{eqnarray}\label{components}
 {D}_{\theta}^{\prime}& =&-i\hbar \left(\frac{1}{R}\frac{\partial}{\partial \theta}-\frac{ie}{\hbar}A_{\theta} \right)\nonumber\\
{D}_{z}^{\prime}& =&-i\hbar \left(\frac{\partial}{\partial z}-\frac{ie}{\hbar}A_{z} \right)
\end{eqnarray}
Then, $H^{cyl}$ can be expressed as
\begin{eqnarray}\label{(H cylinder final}
H^{cyl}&=&\frac{-\hbar^2}{2m}(\hat{\theta}D_\theta^{\prime}+\hat{z}D_z^{\prime})^2-\frac{\hbar^2}{8mR^2}-\frac{e\hbar}{2m}\vc{\sigma}\cdot\vc{B}\nonumber\\
&=&\frac{-\hbar^2}{2m}{\vc{D}}^{\prime}.{\vc{D}}^{\prime}-\frac{\hbar^2}{8mR^2}-\frac{e\hbar}{2m}\vc{\sigma}\cdot\vc{B}
\end{eqnarray}
This form of $H^{cyl}$ guarantees its gauge-covariance  (see the appendix). The construction of the gauge-covariant Hermitian Hamiltonian on the surface of a sphere proceeds in exactly the same way. The Hamiltonian, Eq. (\ref{Pauli equation}), expressed in spherical coordinates reads:
\begin{eqnarray}\label{sphere}
H&=&\frac{-\hbar^2}{2m}\nabla^2 +\frac{i\hbar e}{m}\left(A_{r}\frac{\partial}{\partial r}+\frac{1}{r} A_{\theta}\frac{\partial}{\partial \theta}+\frac{1}{r\sin\theta} A_{\phi}\frac{\partial}{\partial\phi}\right)\nonumber\\&+&\frac{i\hbar e}{2m}\left(\frac{1}{r^2}\frac{\partial(r^{2}A_{r})}{\partial r}+\frac{1}{r\sin\theta}\frac{\partial(\sin\theta A_{\theta})}{\partial \theta}+\frac{1}{r\sin\theta}\frac{\partial A_{\phi}}{\partial\phi} \right)+ \frac{e^2}{2m}\vc{A}.\vc{A}-\frac{e\hbar}{2m}\vc{\sigma}\cdot\vc{B}
\end{eqnarray}
where $ \nabla^2$ is the Laplacian in spherical coordinates; Eq(\ref{laplacian spherical}) . We leave it as an exercise to the reader to check that the pragmatic approach to construct the Hamiltonian on the surface of a sphere leads to a non-Hermitian Hamiltonian and move on to construct the Hermitian one.We first identify the radial component of the gauge-covariant kinematical momentum in spherical coordinates $$\Pi_{r}=-i\hbar (p_{r}-\frac{ie}{\hbar}eA_{r})$$
with $p_{r}$ now, being the canonical Hermitian radial momentum given by Eq.(\ref{hermitian p sphere}). We have for the above $\Pi_{r} $:
 \begin{eqnarray}\label{(k momentum)^2,sphere}
\frac{ {\Pi_{r}}^2}{2m} &=&\frac{-\hbar^2}{2m}\left(\frac{1}{r^2}\frac{\partial}{\partial r}(r^2\frac{\partial}{\partial r})\right)+\frac{i\hbar e}{m} A_{r}\frac{\partial}{\partial r}\nonumber \\&+&\frac{i\hbar e}{2m}\left(\frac{2 A_{r}}{r}+\frac{\partial A_{r}}{\partial r} \right)
 \end{eqnarray}
 Thus, the Hamiltonian , Eq. (\ref{sphere}), expressed in terms of $ \frac{ {\Pi_{r}}^2}{2m}$ reads:
 \begin{eqnarray}\label{sphere prime}
H&=&\frac{{\Pi_{r}}^2}{2m} +\frac{-\hbar^2}{2m}\left(
\frac{1}{r^2\sin\theta}\frac{\partial}{\partial\theta}(\sin\theta\frac{\partial}{\partial\theta})+\frac{1}{r^2\sin^2\theta}\frac{\partial^2}{\partial\phi^2}\right) +\frac{i\hbar e}{m}\left(\frac{1}{r} A_{\theta}\frac{\partial}{\partial \theta}+\frac{1}{r\sin\theta} A_{\phi}\frac{\partial}{\partial\phi}\right)\nonumber\\&+&\frac{i\hbar e}{2m}\left(\frac{1}{r\sin\theta}\frac{\partial(\sin\theta A_{\theta})}{\partial \theta}+\frac{1}{r\sin\theta}\frac{\partial A_{\phi}}{\partial\phi} \right)+ \frac{e^2}{2m}(A_{\theta}^{2}+A_{\phi}^{2})-\frac{e\hbar}{2m}\vc{\sigma}\cdot\vc{B}
\end{eqnarray}
when $\Pi_{r} \rightarrow 0$, and $r \rightarrow  R$, the above Hamiltonian expressed in terms of covariant derivatives becomes:
\begin{eqnarray}\label{sphere prime final}
H^{sph}&=&\frac{-\hbar^2}{2m}\left(\hat \theta(\frac{1}{R}\frac{\partial}{\partial \theta}-\frac{ie}{\hbar}A_{\theta}) +\hat \phi(\frac{1}{R\sin\theta}\frac{\partial}{\partial \phi}-\frac{ie}{\hbar}A_{\phi})\right)^2-\frac{e\hbar}{2m}\vc{\sigma}\cdot\vc{B}\nonumber\\&=&\frac{-\hbar^2}{2m}(\hat{\theta}D_\theta^{\prime}+\hat{\phi}D_\phi^{\prime})^2-\frac{e\hbar}{2m}\vc{\sigma}\cdot\vc{B}\nonumber\\&=&\frac{-\hbar^2}{2m}{\vc{D}}^{\prime}.{\vc{D}}^{\prime}-\frac{e\hbar}{2m}\vc{\sigma}\cdot\vc{B}
\end{eqnarray}
where again we have defined the on-the-surface covariant derivatives for the sphere as:
$$D_\theta^{\prime}=-i\hbar \left(\frac{1}{R}\frac{\partial}{\partial \theta}-\frac{ie}{\hbar}A_{\theta}\right) $$
$$D_\phi^{\prime}=-i\hbar\left(\frac{1}{R\sin\phi}\frac{\partial}{\partial \phi}-\frac{ie}{\hbar}A_{\phi} \right) $$
and one has to remember that $ \frac{\partial\hat{\phi}}{\partial \theta}=0$, $ \frac{\partial\hat{\theta}}{\partial \theta}=-\hat{r}$, $ \frac{\partial\hat{\phi}}{\partial \phi}=-\cos\theta \hat{\theta}-\sin\theta \hat{r}  $,  $ \frac{\partial\hat{\theta}}{\partial \phi}=\cos\theta \hat{\phi}$.
\section{conclusions}
Hermitian Schr\"{o}dinger Hamiltonian for a spin-less particle  as well as  Hermitian and gauge-covariant Pauli Hamiltonian for a spin one-half particle in a magnetic field  on the surfaces of a cylinder and a sphere has been constructed. The methodology is based on starting with a Hamiltonian in the 3D space and achieving confinement by assuming a strong confining radial potential that pins the particle to the surface, thus freezing the radial degrees of freedom. In dropping these from the 3D Hamiltonian to construct the surface one, we have demonstrated how to correctly construct the  Hermitian radial momentum operator that need to be dropped. Dropping blindly only the radial derivative from the laplacian operator has been shown to lead to non-Hermitian surface Hamiltonian on the surface. In the presence of a magnetic field, we have demonstrated that it is the radial component of the gauge-covariant kinematical momentum that should be identified with the physical radial degree of freedom to be dropped; leading to a Hermitian and at the same time gauge-covariant surface Hamiltonian. \\
While the treatment in this work has been restricted to a cylinder and a sphere for clarity and accessibility,the reader interested in considering he application of the procedure to other  surfaces is referred to the general treatment in  \cite{shikakhwa and chair2}  .

\section*{Appendix1}

It was  shown about four decades ago \cite{Koppe,Costa} that when a particle is confined to a two-dimensional surface embedded in 3-dimensional space,the quantum kinetic energy of the particle acquires an extra geometrical term and reads:
$$ -\frac{\hbar^2}{2m}\nabla^2 -\frac{\hbar^2}{2m}( M^2-K)$$
where the quantities  $M$ and $K$ are, respectively,  the mean and the Gaussian curvatures of the surface; which are two standard quantities in the theory of surfaces  (see \cite{Grinfeld}, for example). The mean curvature is given by  the arithmetic mean of two more geometrical quantities; the principal curvatures of the surface, $ \kappa_{1}$, $\kappa_{2} $  $$ M=\frac{\kappa_{1}+\kappa_{2}}{2},$$ $ \kappa_{1}$, $\kappa_{2} $ are the maximum and minimum values of the normal curvatures of a surface. The normal curvature  can be intuitively thought of as a quantity measuring the bending of a surface towards or away from the normal to the surface ( a simple discussion of these concepts that does not require a prior knowledge of differential geometry can be found in the course notes \cite{Jeff}). The Gaussian curvature is also given in terms of $ \kappa_{1}$, $\kappa_{2} $, but as a product, $$ K=\kappa_{1}\kappa_{2}$$ These formulas show that the mean curvature $M$ is non-vanishing if one of the principal curvatures is non- zero, it may happen that $\kappa_{1}=-\kappa_{2}$ in this case $M=0$. For the Gaussian curvature, $K\neq 0$, if both principal curvatures are non-vanishing. For a cylinder of radius $R$, $ \kappa_{1}=-\frac{1}{R}$, $\kappa_{2}=0 $ \cite{Grinfeld} giving $M=-\frac{1}{R}$ and $K=0$, thus we get  the GKE given in the text $-\frac{\hbar^2}{2m}( M^2-K)=-\frac{\hbar^2}{8mR^2}$. For a sphere, on the other hand, we have $\kappa_{1}=\kappa_{2}= -\frac{1}{R}$ giving zero GKE as in the text.

\section*{Appendix2}

\ Consider a Hamiltonian $H(\vc{A})$ that is a function of a the vector potential $ \vc{A}$, with the time-independent Schrodinger equation being
\begin{equation}\label{Sch1}
H(\vc{A})\Psi=E\Psi
\end{equation}
If under the simultaneous transformations $ \vc{A}\rightarrow \vc{A'}=\vc{A}+\vc{\nabla} \lambda$, $ \Psi\rightarrow \Psi^{\prime}=U\Psi=\exp({\frac{ie}{\hbar}}\lambda)\Psi$ for $\lambda$ an arbitrary differentiable function of position and possibly time, the above Schr\"{o}dinger equation transforms to:
\begin{equation}\label{Sch2}
H(\vc{A'})\Psi^{\prime}=E\Psi^{\prime}
\end{equation}
we say that the Schrodinger equation is gauge-covariant. Now, it can be easily seen that  if $H(\vc{A'})=U H(\vc{A}) U^{-1} $, i.e, if the gauge transformation   $\vc{A}\rightarrow \vc{A'}=\vc{A}+\vc{\nabla} \lambda$ represents  a unitary transformation of $ H(\vc{A})$, then the Schrodinger equation is gauge-covariant. Simply (recall that $U^{-1}U=UU^{-1}=I$, the identity )
$$H(\vc{A'})\Psi^{\prime}=U H(\vc{A  }) U^{-1}U\Psi=U H(\vc{A  })\Psi=E\Psi^{\prime} $$
Now, if  $H(\vc{A})$ is of the form: $H(\vc{A  })=-\frac{\hbar^{2}}{2m}\vc{D}.\vc{D} +  \dots$, (where dots include either gauge-invariant; $\vc{B}$ for example, or a vector potential-independent terms), then $H(\vc{A  }) $ is indeed gauge-covariant. This is so, because
\begin{eqnarray}\label{gauge inv}
U\vc{D} U^{-1}&=& U\left( \vc{\nabla}-\frac{ie}{\hbar}\vc{A}\right) U^{-1}\nonumber\\&=&U\left( (\vc{\nabla} U^{-1})+ U^{-1}\vc{\nabla}-U^{-1}(\frac{ie}{\hbar}\vc{A})\right)\nonumber\\&=&UU^{-1}\left( \vc{\nabla}-\frac{ie}{\hbar}(\vc{A}+\vc{\nabla}\lambda)\right)=\vc{D}(\vc A^{\prime})
\end{eqnarray}
Therefore,
$$ U \vc{D}(\vc{A}). \vc{D}(\vc{A})U^{-1}=U\vc{D}(\vc{A})U^{-1}.U\vc{D}(\vc{A})U^{-1}=\vc{D}(\vc{A}^{\prime}).\vc{D}(\vc{A}^{\prime})$$

\end{document}